\documentstyle[aps,twocolumn]{revtex}
\begin{document}
\draft
\title{Purely perturbative Boltzmann equation for hot non-Abelian gauge theories}
\author{Yoshitaka Hatta}
\address{Department of Physics, Kyoto University, Kyoto 606-8502 Japan}
\maketitle
\begin{abstract}
In the perturbation theory, trasnport phenomena in hot non-Abelian gauge theories like QCD are often plagued with infrared singularities or nonperturbative effects.
 We show, in the context of the Kadanoff $\&$ Baym formalism, that there are certain nonequilibrium processes which are free from such difficulties.  
For these processes, due to an interplay between the macroscopic and microscopic physics, characteristic time scale (the {\it mesoscale}) naturally enters as an infrared cutoff and purely perturbative description by the Boltzmann equation is valid. 
\end{abstract}

Recently, there have been considerable interests in the transport phenomena in hot gauge theories \cite{blaizot2,bodeker,arnold5,boyanovsky,blaizot4} motivated by the physics of the quark-gluon plasma (QGP) \cite{hwa} and the baryon number violation in the early universe \cite{rubakov}. The standard approach is to derive the Boltzmann equation and obtain various transport coefficients from it. 
However, in this direction one often encounters logarithmic singularities due to the exchange of very soft gluons. Such singularities are typically summarized in the following integral
\begin{eqnarray}
\int^{gT}dq \frac{1}{q},\label{red}
\end{eqnarray}
where $g$ is the gauge coupling constant and $T$ is the temperature.
The upper limit $gT$ is needed to be consistent with the hard thermal loop approximation\cite{braaten} leading to the integrand $1/q$. Due to this logarithmic singularity, some quantities associated with nonequilibrium phenomena, e.g., the damping rate of thermal gluons and fermions, the color relaxation rate \cite{gyulassy,heiselberg} are divergent. Henceforth we call these quantites transport coefficients, although it is an abuse of the nomenclature. In fact, they are {\it superficially} divergent. This is because the integrand $1/q$ is valid only when $q$ is larger than some nonperturbative scale $\mu$.  So we must cut off the integral at this scale anyway and further analyses require nonperturbative considerations.
 $\mu$ is considered to be of order $g^2T$ up to a possible logarithmic factor of $g$\cite{bodeker}.

The above situation seems to indicate that even in the lowest order of perturbation theory one cannot be free of the nonperturbative effects.
In this paper we will show that this is not always the case at least in some physically important processes like the damping of thermally excited particles.

There are several derivations of the Boltzmann equation. Here we follow one by Blaizot $\&$ Iancu \cite{blaizot2} based on the Kadanoff $\&$ Baym formalism \cite{kadanoff} because, as we will see, it is well suited for our purpose. In this formalism, the Boltzmann equation for the pure SU($N_c$) plasma describes the slow variation of the Wigner transform $G(k,X=\frac{x+y}{2})$ of the hard (containing spatial momenta of order $T$) gluon two point function $G(x,y)$ 
\begin{eqnarray}
2[k\cdot D_X, \delta \acute{G}(k, X)]-2gk^{\alpha}F_{\alpha \beta}(X)\partial^{\beta}_kG_{eq}(k)\nonumber \\
=-\Gamma(k)\delta \acute{G}(k, X)+\delta \acute{\Sigma}^>G_{eq}^<-\delta\acute{\Sigma}^<G_{eq}^>\label{col},
\end{eqnarray} 
\begin{eqnarray}
\Gamma(k)&=&\Sigma_{eq}^<(k)-\Sigma_{eq}^>(k),
\end{eqnarray}
where $\delta \acute{G}\equiv \delta \acute{G}^>=\delta \acute{G}^<$ and $\delta \acute{\Sigma}^{\stackrel{>}{<}}$ are the deviations from equilibrium of the covariantized gluon propagator \cite{blaizot2} and the Keldysh components of the gluon self energy transversely projected, respectively. The covariant derivative $D_X^{\mu}$ and the field strength $F^{\mu\nu}(X)$ are constructed from the supersoft (containing spatial momenta of order $g^2T$ ) mean field $A^{\mu}(X)$. 

Let us consider the problem of the damping of a hard gluon using (\ref{col}). To simplify (\ref{col}) one usually employs the {\it Kadanoff $\&$ Baym ansatz} which assumes that the nonequilibrium Green's functions are proportional to the spectral function. Furthermore, one assumes
\begin{eqnarray}
\acute G^>(k,X)-\acute G^<(k,X)\equiv \rho(k,X) \simeq 2\pi \epsilon (k_0)\delta (k^2),\label{quasi}
\end{eqnarray}
where $\epsilon (k_0)$ is the sign function. This is known as the {\it quasiparticle approximation}. (Actually, this approximation is required because when we have written down (\ref{col}) we dropped terms which produce the width in the spectral function \cite{kadanoff}. Note that we have already utilized the fact that $\delta \acute{G}^>=\delta \acute{G}^<$ in (\ref{col}), which is a consequence of the quasiparticle approximation.) Thus, one ends up with the following structure:
\begin{eqnarray}
& & \delta \acute{G}(k,X)\equiv \rho (k,X)\delta N(k,X)\nonumber \\ 
&\simeq& 2\pi \delta(k^2)[\theta(k_0)\delta N({\bf k},X)+\theta(-k_0) \delta N(-{\bf k},X)]\label{an}.
\end{eqnarray}
Dropping the mean field $A^{\mu}$ and neglecting the deviations from the equilibrium value other than the mode with wavevector ${\bf k}$, that is, neglecting $\delta \acute{\Sigma}^{\stackrel{>}{<}}$, we obtain 
\begin{eqnarray}
\frac{d}{dX_0}\delta N({\bf k},X_0)\simeq-2\gamma \delta N({\bf k},X_0),
\end{eqnarray}
where $\gamma\equiv \frac{1}{4k}\Gamma(k_0=k)$ is the damping rate of the hard gluon. It is well known that this quantity is divergent because of the long range nature of the color magnetic force. To see the divergence, we calculate it with 4-vector $k^{\mu}$ slightly off-shell to obtain
\begin{eqnarray}
\Gamma(k)\simeq 4\alpha kTN_c\ln \frac{\omega_p}{|k_0-k|},\ \ k=|{\bf k}|\label{one}
\end{eqnarray}
where $\omega_p$ is the plasma frequency and $\alpha=\frac{g^2}{4\pi}$. We see that the product of $\Gamma(k)$ and the on-shell delta function (\ref{an}) appearing in (\ref{col}) is clearly ill-defined. 

To avoid such a singularity, the first thing one can think of is to include the width to the spectral function. Perturbatively, the delta function approximation is quite a good approximation for a hard gluon. Nevertheless we try to include loop corrections. To the one loop order, we get the plasmon pole. $\Gamma(k)$ is logarithmically divergent at the pole in the same manner as the bare case. In addition, we don't know how to handle with the Landau damping spectrum. To obtain the width, we must proceed to the next order. But the width is essentially the same as the damping rate which is divergent. So this approach is not practical besides the technical complexity.

Another approach is to consider the finite time effect \cite{hatta}. (See also \cite{blaizot4,boyanovsky}.)
Notice that the delta function appears as the Fourier transform of the propagators: The free and equilibrium propagators read
\begin{eqnarray}
G_{\bf k}^>(t,t')& &\nonumber \\
=\frac{1}{2k}\{(1&+&N(k))e^{-ik(t-t')}-(1+N(-k))e^{ik(t-t')}\},\nonumber\\
G_{\bf k}^<(t,t')&=&\frac{1}{2k}\{N(k)e^{-ik(t-t')}-N(-k)e^{ik(t-t')}\},\label{ggg}
\end{eqnarray}
where $N(k)$ is the Bose distribution function.
Fourier transforming with respect to the relative time {\it from minus infinity to plus infinity}, we obtain
\begin{eqnarray}
G^<(k_0,{\bf k})=\int_{-\infty}^{\infty}d(t-t')e^{ik_0(t-t')}G_{\bf k}^<(t-t') \nonumber \\
=2\pi \epsilon(k_0)\delta(k^2)N(k_0).
\end{eqnarray}
Now we perform a modification needed in the presence of infrared singularities.
Consider a situation that at time $t=0$ we add a hard gluon with momenta $k^{\mu}$ to the initially equilibrium plasma. We write the deviation from the equilibrium at $t,t'\ge 0$ as (See (\ref{ggg}).)
 \begin{eqnarray}
\delta \acute{G}_{\bf k}^<(t,t')=\delta \acute{G}_{\bf k}^>(t,t')=\frac{1}{2k}\delta N({\bf k},X_0)e^{-ik(t-t')}.
\end{eqnarray}
Once we have fixed $X_0$, $t$ and $t'$ are constrained as follows 
\begin{eqnarray}
t,t'\ge 0,\ X_0=\frac{t+t'}{2}
 \ \Rightarrow \ \ 2X_0\ge t-t'\ge -2X_0.
\end{eqnarray}
Our point is that we perform the ``Wigner transformation'' in this finite time interval

\begin{eqnarray}
\delta \acute{G}(k,X_0)&=&\int_{-2X_0}^{2X_0}d(t-t')e^{i(k_0-k)(t-t')}\frac{\delta N({\bf k},X_0)}{2k} \nonumber \\
&=&\frac{\delta N({\bf k},X_0)}{2k}\frac{2\sin[2X_0(k_0-k)]}{k_0-k}\label{new}.
\end{eqnarray}
Thus the ``spectral function'' acquires the width $\sim 1/X_0$ by the finite time effect.  Note that because $\delta G_{\bf k}^<=\delta G_{\bf k}^>$, (\ref{new}) is {\it compatible} with the quasiparticle approximation (\ref{quasi}). Substituting (\ref{new}) in (\ref{col}), we get after $k_0$-integration
\cite{for}
\begin{eqnarray}
2k\frac{d}{dX_0}\delta N({\bf k},X_0)\simeq-4\alpha kTN_c\ln(2\omega_pX_0)\delta N({\bf k},X_0),\label{f}
\end{eqnarray}
\begin{eqnarray}
\delta N({\bf k},X_0)\simeq\delta N({\bf k},0)\exp\{-2\alpha TN_cX_0\ln(2\omega_p X_0)\}.\label{none}
\end{eqnarray}
We see that the gluon distribution function exhibits the anomalous (non-exponential) damping \cite{takashiba,blaizot}. 
The infrared singularity is cut off by the inverse of the time scale with which we are looking at the system. Our method leading to (\ref{none}) is different from \cite{blaizot4} in that we have included the finite time effect entirely into the spectral function. This is very convenient in the later argument. 

It is clear that (\ref{none}) is valid only for $1/gT \ll X_0\le 1/{\mu}$. This is because (\ref{one}) is only valid when $gT \gg |k_0-k| \ge \mu$ and in the integral leading to (\ref{f}) the region $ |k_0-k| \sim 1/{X_0} $ is important. However, in this region (\ref{none}) is a definite prediction of the {\it perturbation} theory and no nonperturbative effect comes into play. 

What we have done above can be stated in more general terms. 
Consider a nonequilibrium process (not restricted to gauge theories) which has a characteristic (or macroscopic) scale of the temporal variation. Characteristic scales of the spatial variation are less important compared to the time scale in the following argument. (Note that transport processes occur even in spatially homogeneous systems.) We introduce a time scale $\tau$ such that $\tau$ is much smaller than the characteristic time scale and much larger than $1/gT$. Imagine a space-time lattice whose lattice spacings are given by $\tau$ and characteristic spatial scales. (See, e.g.,\cite{ho,niegawa}.) Each cell is specified by the Wigner coodinate $X^{\mu}$. We can ``Fourier transform'' only within a cell to discuss the difference between neighboring cells, that is, to discuss transport phenomena. Usually, scales of microscopic physics are literally much smaller than macroscopic scales so the integration ranges are practically infinite. But in the present case they cannot be sent to infinity. This effect practically smears the spectral functions making otherwise divergent transport coefficients logarithmically dependent on $\tau$.
It is interesting to see that the infrared singularity originating from the {\it spatially} long range color magnetic force is cut off by the time scale. Here we can see an interplay between the macroscopic and the microscopic physics, or the {\it feedback effect} of the macroscopic physics. Namely, the mass singularity coming from the microscopic collision is remedied by the nonequilibrim nature of the bulk system whose evolution in turn is generated by the accumulation of microscopic collisions. 

In (\ref{col}), $k$ is an arbitrary 4-vector, which enables us to study {\it around the mass shell}. Due to the finite time effect particles participating in a collision have certain off-shellness. This effect can be naturally incorporated in the Kadanoff $\&$ Baym formalism \cite{other}. 

The full Boltzmann equation (\ref{col}) becomes \cite{blaizot2}
\begin{eqnarray}
(v\cdot D_X)^{ab}W_b(X,{\bf v})={\bf v}\cdot {\bf E}^a(X)-\gamma\biggl\{W^a(X,{\bf v})\nonumber \\
-\frac{4}{\pi}\int \frac{d\Omega'}{4\pi}\frac{({\bf v}\cdot {\bf v'})^2}{\sqrt{1-({\bf v}\cdot {\bf v'})^2}}W^a(X,{\bf v}')\biggr\},\label{my}
\end{eqnarray}
where $v^{\mu}=(1,\hat{\bf k})$ and $\delta N({\bf k},X)\equiv -gW(X,{\bf v})\frac{dN}{dk}$. $\gamma$ is commonly called the damping rate of the hard gluon, but the logarithmic singularity is cut off by $\tau$, the intermediate time scale between the macroscopic and microscopic physics. If we are interested in the short time evolution of processes with the characteristic time less than or equal to $1/{\mu}$, we can safely use (\ref{my}) which is free of any nonperturbative effects \cite{mag}. 

Note that the ordinary Boltzmann equation for a dilute classical gas can be used only for an intermediate time scale (the {\it mesoscale}) much longer than the collision time and much shorter than the mean free time as was first pointed out by Bogoliubov \cite{bogoliubov}. The situation is analogous here. $1/{\mu}$ turns out to be of the same order as the mean free time of hard gluons. So the {\it purely perturbative Boltzmann equation} proposed here is in accordance with Bogoliubov's spirit and describes intermediate processes leading to the damping of the hard gluon and the color transportation\cite{a}. This is our main conclusion.

 The above consideration cannnot be extended to the situations when the characteristic time scale is larger than $1/{\mu}$.  Beyond this regime, we must cut off the integral by the {\it momentum} cutoff $\mu$ instead of $|k_0-k|$. So we must regard the resulting equation as an effective theory and further analysis requires some nonperturbative methods. 
One of the most interesting examples of this is the baryon number violation in the early universe. It was shown \cite{arnold} that the characteristic time scale of that process is $1/g^4T$ much larger than $1/{\mu}$. The effective theory was constructed by B\"odeker \cite{bodeker} and the hot sphaleron rate has been calculated by Moore on the lattice \cite{moore}. 

In summary, we have shown, in the Kadanoff $\&$ Baym formalism, that the nonequilibrium non-Abelian plasma may exhibit mesoscopic physics which can be described by the purely perturbative Boltzmann equation. Although we do not know phenomenologically interesting processes that have such short characteristic time scales at present, these may be found, for example, in the initial stage of the QGP in future ultrarelativistic heavy ion experiments.   \\  

I would like to thank T. Kunihiro for carefully reading the manuscript and giving useful comments. I also thank A. Niegawa, D. N. Voskresensky and the members of the HD seminar at the Nuclear Theory Group of Kyoto University for helpful discussions.


%
%

%
%


\begin{references}
\bibitem{blaizot2} J. P. Blaizot and E. Iancu, Nucl. Phys. {\bf B557}, 183 (1999); Nucl. Phys. {\bf B570}, 326 (2000).

\bibitem{bodeker} D. B\"odeker, Phys. Lett. B {\bf 426}, 351 (1998); Nucl. Phys. {\bf B599}, 502 (1999); {\bf B566}, 402 (2000).
\bibitem{arnold5} P. Arnold, D. T. Son and L. G. Yaffe, Phys. Rev. {\bf D59}, 105020 (1999).
\bibitem{boyanovsky}
D. Boyanovsky, H. J. de Vega, and S. Y. Wang,
Phys. Rev. D {\bf 61}, 065006 (2000);
S. Y. Wang, D. Boyanovsky, H. J. de Vega and D. S. Lee, Phys. Rev. D {\bf 62}, 105026 (2000).
\bibitem{blaizot4}  J. P. Blaizot and E. Iancu, hep-ph/0101103.
\bibitem{hwa} {\it Quark-Gluon Plasma}, ed. R. Hwa, (Advanced Series on Directions in High Energy Physics, World Scientific 1990); {\it Quark-Gluon Plasma II}, ed. R. Hwa, (Advanced Series on Directions in High Energy Physics, World Scientific 1995).
\bibitem{rubakov} V. A. Rubakov and M. E. Shaposhnikov, Usp. Fis. Nauk. {\bf 166}, 493 (1996); hep-ph/9603208.

\bibitem{kadanoff}  L. P. Kadanoff and G. Baym, {\it Quantum Statistical Mechanics} (W. A. Benjamin, New York 1962).

\bibitem{braaten} E. Braaten and R. D. Pisarski, Nucl. Phys. {\bf B337} (1990) 569; {\it ibid}. {\bf B339} (1990) 310.

\bibitem{hatta} Y. Hatta, poster presented at YITP workshop ``Fundamental Problem and Applications of Quantum Field Theory'' at Yukawa Institute for Theoretical Physics (Dec. 20-22, 2000); Y. Hatta, master thesis (Kyoto University, 2001).

\bibitem{for}We have used the following formula
\begin{eqnarray}
\int_{-\infty}^{\infty} \frac{\sin kx}{x}\ln |x|dx=
-\pi \mbox{sgn}(k)(\ln|k|+\gamma_E)\nonumber,
\end{eqnarray}
($\gamma_E$ is the Euler constant) in the r.h.s., while in the l.h.s. we have used the delta function. We may have been sloppy about this point. But our point is that the broadening of the spectral function regularizes the singularity associated with {\it collisions}. So the drift term representing the free motion can be treated with the delta function as before. 
\bibitem{takashiba} K. Takashiba, Int. J. Mod. Phys. A {\bf 11}, 2309 (1996).

\bibitem {blaizot} J. P. Blaizot and E. Iancu, Phys. Rev. Lett. {\bf 76}, 3080 (1996); Phys. Rev. D {\bf 55}, 973 (1997).


\bibitem{ho} E. Calzetta and B. L. Hu, Phys. Rev. D {\bf 37}, 2878 (1988). 
\bibitem{niegawa} A. Niegawa, Prog. Theor. Phys. {\bf 102}, 1 (1999); hep-th/9810043.

\bibitem{other} There are other approaches to the trasnport phenomena in field theories which do not introduce the Wigner coordinate and directly deal with the microscopic time (see, e.g., \cite{boyanovsky}). We believe that our results can be incorporated in these more fundamental approaches which do not rely on such an ansatz like (\ref{quasi}).


\bibitem{mag} In cases where (\ref{my}) with the cutoff $\tau$ can be reliable, it is clear that the possible {\it magnetic mass} \cite{linde,pisarski} of order $g^2T$ plays no role even if it exists. In cases where we must use $\mu$ as a cutoff, the magnetic mass will enter the successive nonperturbative analysis (see the following paragraph). In both cases the magnetic mass doesn't appear explicitly in the Boltzmann equation.

\bibitem{bogoliubov} N. N. Bogoliubov, in {\it Studies in Statistical Mechanics,} vol. 1, edited by J. de Boer and G. E. Uhlenbeck (North-Holland, Amsteredam, 1962).

\bibitem{a} When we linearize the Boltzmann equation with respect to the deviation from the (local) equilibrium distribution function, we are supposed to consider phenomena with the characteristic time much longer than the mear free time. However, the relaxation time for color fluctuations is of the same order as the mean free time. Thus, the equation for this mode may still be thought as a kinetic equation even after the linearization. 
\bibitem{arnold} P. Arnold, D. Son, L. G. Yaffe, Phys. Rev. D {\bf 55}, 6264 (1997).
\bibitem{moore} G. D. Moore, Nucl. Phys. {\bf B568}, 367 (2000).



\bibitem{gyulassy} A. Selikhov and M. Gyulassy, Phys. Lett. B {\bf 316}, 373 (1993).

\bibitem{heiselberg} H. Heiselberg, Phys. Rev. Lett. {\bf 72}, 3013 (1994).
\bibitem{linde} A. Linde, Phys. Lett. B {\bf 96}, 289 (1980).

\bibitem{pisarski} D. J. Gross, R. D. Pisarski and L. G. Yaffe, Rev. Mod. Phys. {\bf 53}, 43 (1980).


\end{references}
\end{document}